\documentstyle[eqsecnum,prd,aps]{revtex}
\begin{document}
\draft

%
%
\twocolumn[\hsize\textwidth\columnwidth\hsize\csname
@twocolumnfalse\endcsname

\preprint{SUSSEX-AST-95/11-4, IEM-FT-119/95, astro-ph/9511029}
\tighten
\title{Metric perturbations in two-field inflation}
\author{Juan Garc\'\i a--Bellido and David Wands}
\address{Astronomy Centre, School of Mathematical and
Physical Sciences, \\
University of Sussex, Brighton BN1 9QH,~~U.K.}
\date{\today}
\maketitle
\begin{abstract}
We study the metric perturbations produced during inflation in models
with two scalar fields evolving simultaneously.  In particular, we
emphasize how the large-scale curvature perturbation $\zeta$ on fixed
energy density hypersurfaces may not be conserved in general for
multiple field inflation due to the presence of entropy as well as
adiabatic fluctuations.  We show that the usual method of solving the
linearized perturbation equations is equivalent to the recently
proposed analysis of Sasaki and Stewart in terms of the perturbed
expansion along neighboring trajectories in field-space.  In the case
of a separable potential it is possible to compute in the slow-roll
approximation the spectrum of density perturbations and gravitational
waves at the end of inflation. In general there is an inequality
between the ratio of tensor to scalar perturbations and the tilt of
the gravitational wave spectrum, which becomes an equality when only
adiabatic perturbations are possible and $\zeta$ is conserved.
\end{abstract}

\pacs{98.80.Cq \hspace*{2.1cm} Preprint \ SUSSEX-AST 95/11-4, \,
IEM-FT-119/95, \, astro-ph/9511029}

\vskip2pc]
\section{Introduction}

Inflation is the only known mechanism that solves the horizon and
homogeneity problems of hot big bang cosmology~\cite{Book}. However,
the main observational prediction of inflationary models is the
spectrum of density and gravitational wave perturbations they produce.
Observations of temperature anisotropies in the microwave background,
strictly speaking only provide an upper limit on the amplitude of such
perturbations, and could in principle be produced by some other source
of inhomogeneities. Nonetheless, the apparently Gaussian and nearly
scale-invariant nature of the observed perturbations are natural
properties of those produced by quantum fluctuations of the inflaton
field during inflation. If inflation is indeed responsible for the
observed anisotropies of the microwave background and the initial
curvature perturbations from which galaxies formed, then the amplitude
over a limited range is already constrained by
observations~\cite{LL93,tiltvalue,Gorski94}. In future both the range
and precision of these constraints promise to improve considerably and
so it will be increasingly important to fully understand the
predictions made by the inflationary paradigm and the robustness of
these predictions.

Until comparatively recently it was often stated that inflation
predicts a scale-invariant Harrison-Zel'dovich spectrum of density
perturbations with a negligible amplitude of gravitational waves. In
fact, both the tilt of the spectrum and the relative contribution of
gravitational waves to the microwave background anisotropies are
model-dependent quantities~\cite{LL93}. In the conventional model of
inflation driven by the potential energy density of a single
slowly-rolling scalar field, the tilt and the ratio of tensor
(gravitational wave) to scalar (density) perturbations in the
microwave background can be determined by the slow-roll parameters
which describe the slope and curvature of the potential as the
perturbations cross outside the horizon during
inflation~\cite{recon}. Scalar curvature perturbations $\zeta$
can be understood as originating from quantum fluctuations of the
inflaton field that perturb the time it takes to end
inflation~\cite{GP82}, $\zeta = H\,\delta t=H\,\delta\phi/\dot\phi$.
Our ability to determine the linear perturbation at late times solely
in terms of the parameters at horizon crossing depends on the
constancy of $\zeta$ on scales far outside the horizon. The value of
$\zeta$ when a given comoving scale leaves the horizon during
inflation can then be equated with that at re-entry during the
radiation- or dust-dominated eras.

However, most models of particle physics predict not only one but many
coupled scalar fields and in the presence of additional interacting
fields we must re-evaluate these results. The conservation of $\zeta$
relies on the perturbations being adiabatic during inflation. In the
case of more than one field evolving, there is the possibility of
entropy as well as adiabatic fluctuations during inflation. We assume
that all our scalar fields will eventually decay during reheating and
only adiabatic perturbations remain after inflation. The validity of
this assumption is of course very dependent upon the reheating
mechanism, but we will leave its investigation for future work.

In this paper we will illustrate the wider range of behavior possible
in multiple field inflation from the simultaneous evolution of two
fields. Some previous studies of inflation involving two fields, such
as hybrid~\cite{Hybrid,FirstOrder} or extended~\cite{ExtInf}
inflation, only considered the case where one field evolves during
inflation and the role of the second field is just to end inflation by
a sudden phase transition, so the single field results apply. Double
inflation models~\cite{Double} invoke consecutive periods of inflation
driven by two non-interacting fields. The density perturbation spectra
produced by including an interaction term between these fields was
investigated in Ref.~\cite{inter}. Perturbations in models that
involve two interacting scalar fields have usually been considered in
the context of Brans-Dicke gravity~\cite{ExtChaot,DGL,SY95} or more
general scalar-tensor theories~\cite{MM92,GBW}, where the dilaton is
expected to vary together with the inflaton field during inflation.
However, it is the evolution of the second field rather than its
coupling to the metric tensor that we wish to consider here. Only very
recently have analytic results for general multiple-field inflation
been presented by Sasaki and Stewart~\cite{SS95}.

We will show how to evaluate in the slow-roll approximation the
curvature perturbation at the end of inflation, using two alternative
approaches corresponding to different gauge choices. In order to
calculate the perturbation, we need to explicitly integrate along
classical trajectories, and it turns out that it is only possible if
the potential in separable in the different fields. We find that the
presence of entropy fluctuations modifies the usual results for the
scalar spectra produced by inflation. In particular, we show that the
non-conservation of $\zeta$ far outside the horizon leads to the
violation of the usual consistency relation between the ratio of
tensor to scalar perturbations and the tilt of the gravitational wave
spectrum.

\section{Metric perturbations}

We will consider linear perturbations about a spatially flat
Friedmann-Robertson-Walker (FRW) metric with scale factor $a(t)$.
The most general scalar and tensor metric perturbations can be
written as~\cite{B80,KS84,MFB92}
\begin{eqnarray}\label{pertbn}
ds^2 & = & -\ (1+2A) dt^2 + 2B_{,i}dx^idt +\nonumber \\ [2mm]
& & \ a^2(t) \Big[ (1-2{\cal R}) \delta_{ij}
+ 2E_{,ij} + h_{ij} \Big] dx^i dx^j \,,
\end{eqnarray}
where $A$, $B$, $E$ and ${\cal R}$ are scalar perturbations and
$h_{ij}$ is a transverse traceless tensor perturbation corresponding
to gravitational waves. The perturbations can be decomposed into Fourier
modes (with comoving wavenumber $k$) which can be treated separately
in the linear approximation where they decouple.

However not all the scalar perturbations are physical degrees of
freedom and to remove gauge artifacts we can define gauge invariant
quantities~\cite{B80,KS84,MFB92}
\begin{eqnarray}
&\Phi \equiv A + \left(B - a^2\dot{E}\right)\dot{} \,,\\[2mm]
&\Psi \equiv {\cal R} - H\left(B - a^2\dot{E}\right) \,,
\end{eqnarray}
where $H = \dot a/a$ is the Hubble rate of expansion. Note that these
are equivalent to the metric perturbations $A$ and ${\cal R}$ in the
longitudinal gauge, where $E=B=0$. Moreover, for any perturbations
whose spatial part of the stress energy tensor is diagonal, the
equations of motion require $\Psi=\Phi$~\cite{MFB92}, so considering
only linear perturbations the metric has just one scalar degree of
freedom.

The scalar ${\cal R}$ is the intrinsic curvature perturbation on
hypersurfaces of fixed time $t$, which transforms under a gauge
transformation, $t\to t+\xi^0$ as ${\cal R}\to {\cal R} +
H\,\xi^0$~\cite{MFB92}. It is convenient to evaluate the curvature
perturbation on a hypersurface $\Sigma$ of constant energy density
$\rho$,\footnote{In single-field inflation, this corresponds to a
comoving hypersurface.}  corresponding to the choice of gauge $\xi^0 =
\delta\rho/\dot\rho$~\cite{DM95}. This gives
$\,{\cal R}_\Sigma = {\cal R} + H \,\delta\rho/\dot\rho$.  The
intrinsic three-curvature on this surface is $^{(3)}\!R =
4\,\nabla^2{\cal R}_\Sigma$. Note that if we start from a flat
hypersurface (${\cal R}=0$), we can interpret ${\cal R}_\Sigma$ as
being due to a perturbation in the logarithm of the scale factor (or
number of $e$-foldings $N$) on that hypersurface
\begin{equation}\label{Hdt}
{\cal R}_\Sigma = \delta N =  H \delta t \, ,
\end{equation}
where $\delta t = \delta\rho/\dot\rho$. In particular for inflation
with a single field we can write $\delta t = \delta\phi/\dot\phi$,
which gives the familiar result for the origin of density
perturbations~\cite{GP82}.

We will define a quantity
\begin{equation}\label{ZETA}
\zeta \equiv \
 {\cal R}_\Sigma - \frac{\nabla^2\Phi}{3 \dot{H}}\ = \
 \Phi - {H^2\over\dot{H}} \left( \Phi + H^{-1}\dot\Phi \right) \,,
\end{equation}
written in terms of the gauge invariant metric potential, which
coincides with ${\cal R}_\Sigma$ on scales far outside the horizon
($k\ll aH$).
The time-dependence of $\zeta$ on these super-horizon scales is given
by
\begin{equation}\label{dotzeta}
\dot\zeta \ \simeq \ 3 H \left( {\dot p \over \dot\rho}
  - {\delta p \over \delta\rho} \right) \, \zeta \,,
\end{equation}
where $\delta p$ and $\delta\rho$ are the perturbations in the
pressure and energy density on spatially flat hypersurfaces. As we can
see, Eq.~(\ref{dotzeta}) vanishes for adiabatic perturbations. This is
the case for inflation with a single field, so we can evaluate the
curvature perturbation at late times by equating it with that at
horizon crossing. In fact, $\zeta$ continues to be conserved on
super-horizon scales during the radiation- and dust-dominated eras,
and therefore, we can directly equate the curvature perturbation when
it left the horizon during inflation with that at re-entry.

The expression in Eq.~(\ref{ZETA}) allows one to evaluate $\zeta$ once
we know $\Phi$. To that end, one can integrate the coupled
perturbation equations for $\Phi$ and the scalar fields in a
particular gauge, see for instance~Ref.\cite{SY95,GBW}. This allows us
to give an expression for $\zeta$ at the end of inflation in terms of
the field fluctuations at horizon crossing. Alternatively, one could
find $\zeta$ by evaluating the perturbation in the number of
$e$-foldings, $\delta N$, given in Eq.~(\ref{Hdt}),
\begin{equation}\label{dNzeta}
\zeta \simeq {\cal R}_\Sigma = \delta N \,,
\end{equation}
as proposed recently by Sasaki and Stewart~\cite{SS95}. On spatially
flat hypersurfaces the perturbed fields on super-horizon scales
effectively obey the same equations of motion as the homogeneous
background fields, see Eq.(\ref{eqdot}). Thus $\delta N$ is the
difference in the number of $e$-foldings along neighboring classical
trajectories in phase-space. In order to evaluate $\delta N$ at the
end of inflation in terms of field fluctuations at horizon crossing,
we have to integrate the background equations of motion, not only
along a single trajectory but also along the perturbed
trajectories. We will show that this does indeed yield the same
results as those obtained by directly solving the evolution of the
metric perturbations in the longitudinal gauge. The difference between
the two approaches is just a choice of gauge.

In practice, whichever method one adopts, one needs to know not only
the initial perturbation but also its integrated effect along the
subsequent trajectory. In the following sections we shall show how the
perturbations can be evaluated in specific models. However it is
important to remember that $\zeta$ only remains a conserved quantity
thereafter if we can treat the end of inflation as a transition at
fixed energy density and the subsequent evolution is adiabatic. If we
wish to match metric perturbations at the end of inflation, across a
hypersurface of fixed energy density, we must match $\zeta$ on large
scales~\cite{DM95}. While we expect this to be true in single field
models of inflation, it is a much more complicated issue in two-field
inflation, since it depends upon the dynamics of reheating. Such an
involved issue deserves further attention and is beyond the scope of
this paper.

The attractive feature of Sasaki and Stewart's approach is that
Eq.~(\ref{dNzeta}) is a purely geometrical result, independent of the
matter content (subject only to the condition $\Psi=\Phi$) and could
in principle also be applied to calculate $\zeta$ on surfaces long
after the end of inflation. However evaluating $\delta N$ in the
radiation-dominated era once again requires a quantitative
understanding of reheating along different trajectories. In what
follows we will restrict ourselves to a calculation of the curvature
perturbation at the end of inflation.

\section{Two-field models of inflation}
\label{sectstt}

In this section we will consider a model with two scalar fields,
described by the action
\begin{eqnarray}\label{action}
S & = & \int d^4x \sqrt g \left[ {1\over2\kappa^2}\, R
	- {1\over2} g^{ab} \phi_{,a} \phi_{,b} \right.
\nonumber \\ [2mm]
& & \ \ \left. -\, {1\over2} e^{-2h(\phi)}\, g^{ab} \sigma_{,a} \sigma_{,b}
	- W(\phi,\sigma) \right] \, ,
\end{eqnarray}
where $R$ is the usual Ricci curvature scalar and $\kappa^2 = 8\pi
G$. If $h=0$ then the fields have standard kinetic terms, but we have
also allowed for the possibility that the $\sigma$ kinetic term has a
$\phi$-dependent pre-factor as would come from a conformal
transformation of a theory with a non-minimally coupled $\phi$ field.
Such an action might arise, for instance, in the Einstein
frame\footnote{The original Jordan frame, in which $\sigma$ is
minimally coupled, is related to the Einstein frame used here via the
conformal transformation $\bar g_{ab} = e^{2h(\phi)}\,g_{ab}$.} of
general scalar-tensor gravity theories~\cite{GBW}, in which case
$W(\phi,\sigma)=e^{-4h(\phi)}V(\sigma)$.

The field equations for the fields $\sigma$ and $\phi$
in a spatially flat FRW metric are then
\begin{eqnarray}\label{eqdot}
\ddot\sigma + 3H\dot\sigma &=&
 	- \,e^{2h(\phi)}\ {\partial W \over \partial\sigma} \,
	+ 2 h'(\phi)\,\dot\phi\,\dot\sigma \ , \nonumber\\[2mm]
\ddot\phi + 3H\dot\phi &=& - \,{\partial W \over \partial \phi}
	- h'(\phi)\,e^{-2h(\phi)}\,\dot\sigma^2\ , \\[2mm]
\dot H &=& -\ {\kappa^2\over2} \left( \dot\phi^2
	+ e^{-2h(\phi)}\,\dot{\sigma}^2 \right)\ , \nonumber
\end{eqnarray}
and the Hamiltonian constraint is
\begin{equation}\label{econ}
H^2 = {\kappa^2 \over 6} \left[\dot\phi^2 + e^{-2h(\phi)}\,
	\dot\sigma^2 + 2 W(\phi,\sigma)\right] \ .
\end{equation}
The condition for inflation to occur $\ |\dot H| <
H^2\ $ is thus, see Eqs.~(\ref{eqdot}) and (\ref{econ}),
\begin{equation}\label{end}
\dot\phi^2 + e^{-2h}\,\dot\sigma^2 < \, W(\phi,\sigma)\ .
\end{equation}

In calculating perturbations on a comoving scale $k$, we shall see
that an important quantity is the number of $e$-folds from the end of
inflation when that scale crossed outside the horizon ($k = a_*H_*$),
\begin{equation}\label{efold}
N = - \int_e^* H \ dt \,.
\end{equation}
Our present horizon crossed outside the Hubble scale about 50 to 60
$e$-foldings before the end of inflation. The precise number
depends logarithmically on the energy scale during inflation and the
efficiency of reheating, and so is weakly model-dependent.

We can derive an exact expression for the time dependence of $\zeta$
using the linearly perturbed field equations \cite{GBW}
\begin{equation}\label{DZETA}
\dot\zeta = -\, {H\over\dot H}\, \nabla^2\Phi +
 H \left({\delta\phi\over\dot\phi} -
          {\delta\sigma\over\dot\sigma}\right) \,\Upsilon\ ,
\end{equation}
where
\begin{equation}\label{Upsilon}
\Upsilon = {1\over2}
 {d\over dt} \left( {\dot\phi^2 - e^{-2h}\dot\sigma^2
  \over \dot\phi^2 + e^{-2h}\dot\sigma^2} \right)
 + h' \dot\phi \left( {e^{-2h}\dot\sigma^2
   \over \dot\phi^2 + e^{-2h}\dot\sigma^2} \right)^2 \,.
\end{equation}
Note that although $\delta\phi$ and $\delta\sigma$ are gauge dependent
quantities, their combination in Eq.~(\ref{DZETA}) is gauge
invariant. If only one field is evolving ($\dot\sigma=0$ or
$\dot\phi=0$), we see that $\Upsilon = 0$ and $\zeta$ is conserved on
large scales ($k\ll aH$), thus recovering the well known
result~\cite{Lyth85}. However, if $\Upsilon\neq0$ and both fields are
evolving, $\zeta$ only remains constant for perturbations along the
classical trajectory ($\delta\sigma/\delta\phi =
\dot\sigma/\dot\phi$). The first term on the right-hand-side of
Eq.~(\ref{Upsilon}) is present whenever two fields are evolving. The
second term is due solely to the presence of the non-standard kinetic
term for $\sigma$ and vanishes when $h=$ constant. It represents the
frictional damping of the $\sigma$ field by $h(\phi)$.

\subsection{Slow-roll trajectories}

To make further progress we will work in the slow-roll approximation
in both scalar fields.  In principle this is not necessary for
inflation to occur: one of the fields might roll quickly to the
minimum of its potential and then the problem reduces to single field
inflation. Models of hybrid inflation~\cite{Hybrid} or other models of
first-order inflation~\cite{ExtInf,FirstOrder} provide examples where
more than one field is present but only one field slow-rolls during
inflation.  However, here we wish to consider the case in which both
fields slow-roll. The slow-roll approximation amounts to reducing the
full field equations~(\ref{eqdot}) to first-order equations,
\begin{eqnarray}
3H^2 &\simeq& \ \kappa^2\,W(\phi,\sigma) \ ,\\[2mm]
\dot\sigma &\simeq& -\ e^{2h(\phi)}\,
 {\partial\ln W\over \partial\sigma}
 \, {H\over\kappa^2}\ ,\\[2mm]
\dot\phi &\simeq& -\ {\partial\ln W\over \partial\phi}
 \, {H\over\kappa^2}\ .
\label{SLOW}
\end{eqnarray}
This approximation reduces the effective four-dimen\-sional phase-\-space
during inflation down to the two-dimensional field space
$\{\phi,\sigma\}$.

This leads to a crucial difference between single-field inflation and
inflation with two or more fields. In single-field inflation the
slow-roll solution forms a one-dimensional phase-space, i.e., there is
a unique trajectory. The end of inflation corresponds to a fixed value
of the field and any comoving scale which crossed the horizon $N$
$e$-foldings before the end of inflation also corresponds to a unique
value which may be calculated from Eq.~(\ref{efold}). In two-field
inflation the slow-roll approximation leaves a two-dimensional
phase-space. Classical trajectories during inflation in this field
space correspond to lines which are no longer unique. In particular,
the end of inflation will in general be described by a one-dimensional
line in this phase-space rather than a single point, as will the locus
of a given number of $e$-foldings from the end of inflation.

To calculate $\zeta$ using the formalism of Sasaki and Stewart we need
to know how the dependence of the number of e-foldings from the end of
inflation upon the perturbations in the fields $\phi$ and $\sigma$. In
the single field case this can only amount to a perturbation along the
classical trajectory, $\delta t=\delta\phi/\dot\phi$, due to the
equations of motion~\cite{GP82}.  However, in the two-dimensional
field space, perturbations will in general move the fields onto a
different trajectory with a different end point of inflation, except
when the perturbations happen to be adiabatic.

Therefore, it may no longer be sufficient to use the familiar
relations along a given trajectory $\gamma$,
\begin{equation}
\left( {\partial N\over\partial\phi} \right)_\gamma
 \simeq - {H \over \dot\phi} \, \hspace{1cm}
\left( {\partial N\over\partial\sigma} \right)_\gamma
 \simeq - {H \over \dot\sigma} \,
\end{equation}
in order to work out the change in the number of $e$-foldings due to
perturbations in the fields. To evaluate $\zeta$ at the end of
inflation using Eq.~(\ref{dNzeta}), we have to allow for variations
away from the classical trajectory.

Fortunately, in the case of a separable potential,
\begin{equation}
W(\phi,\sigma) \, = \, U(\phi)\,V(\sigma) \,,
\end{equation}
we can label the slow-roll trajectories by an integral of motion,
\footnote{One can verify from the slow-roll equations of motion that
$\dot{C}\simeq0$ along any classical trajectory $\gamma$.}
\begin{equation}\label{sol}
C = \kappa^2 \int {V\,d\sigma\over V'(\sigma)} \,
     - \, \kappa^2 \int e^{2h(\phi)} {U\,d\phi\over U'(\phi)} \,,
\end{equation}
which allows us to parametrize motion off the classical trajectory.
We can then substitute the slow-roll equation of motion for $\phi$
into Eq.~(\ref{efold}) to obtain the number of $e$-foldings for a
given value of $\phi_*$ along a trajectory labeled by $C$,
\begin{equation}\label{Nefold}
N (\phi_\ast, \phi_e(C) )
 \simeq \,\kappa^2\,\int_e^* {U\,d\phi\over U'(\phi)}\ .
\end{equation}
This depends on the values of both $\phi_*$ and $\sigma_*$ through the
dependence of $\phi_e$ upon $C(\phi_*,\sigma_*)$.

One should, of course, be cautious about evaluating the perturbation
by integrating along a slow-roll trajectory that is only an
approximation to the full equations of motion.  In the single-field
case the conservation of $\zeta$ is an exact result and does not rely
on the slow-roll approximation. In the examples presented in this
paper we have checked by numerical solutions that the slow-roll
results remain a good approximation for a wide range of parameters
right up until the end of inflation.

\subsection{Scalar perturbations}

Perturbations during inflation are expected to arise from the vacuum
fluctuations of the fields which are stretched by the inflationary
expansion up to super-horizon scales.  In the following we refer to
the gauge invariant scalar field perturbations~\cite{MFB92}, or
equivalently the scalar field perturbations in the longitudinal gauge.
For large values of $k \gg aH$ we can neglect the potential terms in
the perturbed field equations for the scalar fields and they reduce to
those of massless fields. Thus, to lowest order in the slow-roll
parameters, the expectation values of the perturbations as they cross
outside the Hubble radius ($k \simeq a_*H_*$) are given by Gaussian
random variables with $\ e^{-2h_*}\,\langle|\delta\sigma_*|^2\rangle =
H_*^2/2k^3 \ , \ \ \langle|\delta\phi_*|^2\rangle = H_*^2/2k^3
\, ,\ $ where $k$ is the comoving wavenumber. Note that, while the
field $\phi$ has a standard kinetic term in Eq.~(\ref{action}), the
$\sigma$ field does not and so the $\phi$-dependent prefactor must be
included in the expectation value acquired at horizon
crossing~\cite{JGBNuclPhys}. We shall denote the
spectrum of a quantity $A$ by $\ {\cal P}_A (k)
\equiv (4\pi k^3/ (2\pi)^3)\,\langle|A|^2 \rangle\, ,\ $ as defined
in~\cite{LL93}. Thus we have
\begin{eqnarray}
\label{Pdphi}
{\cal P}_{\delta\phi} &\simeq &
\left( {H_\ast \over 2\pi} \right)^2 \,,\\
\label{Pdsigma}
{\cal P}_{\delta\sigma} &\simeq & \,
e^{2h_\ast}\, \left( {H_\ast \over 2\pi} \right)^2 \,.
\end{eqnarray}

For slowly varying, long-wavelength ($k \ll aH$)
modes, to lowest order in the slow-roll parameters, the
perturbation equations can be integrated to give~\cite{GBW},
\begin{eqnarray}
\delta\phi &\simeq&{\displaystyle
\ {1\over\kappa^2}\,{U'(\phi)\over U(\phi)}
\,Q_1 }\,,\\[2mm]
\delta\sigma &\simeq&{\displaystyle
\ {1\over\kappa^2}\,{V'(\sigma)\over
V(\sigma)}\,\left(Q_2 + e^{2h} Q_1 \right) }\ .
\label{DDSIG}
\end{eqnarray}
where $Q_1$ and $Q_2$ are constants of integration. It will be
convenient to define a new constant $Q_3\equiv Q_2 + e^{2h_*} Q_1$, so
that $Q_1$ and $Q_3$ are independent Gaussian random variables whose
values, for a given Fourier mode, are determined by the amplitude of
$\delta\sigma_*$ and $\delta\phi_*$ at horizon-crossing. Thus they
have expectation values
\begin{eqnarray}
&{\displaystyle
{\cal P}_{Q_1} = {\kappa^2\over2\epsilon_\phi^*}\,
\left({H_*\over2\pi}\right)^2 }\label{Q1} \,, \\
&{\displaystyle
{\cal P}_{Q_3} = {\kappa^2\,e^{2h_*}\over2\epsilon_\sigma^*}\,
\left({H_*\over2\pi}\right)^2 }\label{Q3} \ .
\end{eqnarray}

During slow-roll, in the long wavelength limit, the curvature
perturbation on hypersurfaces of constant energy density can be
written as~\cite{MM92,GBLL}
\begin{eqnarray}\label{SRZETA}
\zeta \,& \simeq & H\,{\dot\phi\delta\phi + e^{-2h}\dot\sigma\delta\sigma
\over\dot\phi^2 + e^{-2h}\dot\sigma^2} \ , \nonumber \\[2mm]
& \simeq & \ {\left[ \epsilon_\phi + (e^{2h}-e^{2h_*})\,
\epsilon_\sigma\right]Q_1 + \epsilon_\sigma Q_3 \over \epsilon_\phi
+ e^{2h}\, \epsilon_\sigma} \ ,
\end{eqnarray}
where we have extended the usual definition of the slow-roll parameter
$\epsilon$ for a single field~\cite{LPB94} to
\begin{equation}\label{SRP}
\epsilon_\sigma \equiv {1\over2\kappa^2}
\left({V'(\sigma)\over V(\sigma)} \right)^2 \ , \hspace{1cm}
\epsilon_\phi \equiv {1\over2\kappa^2} \left({U'(\phi)\over U(\phi)}
	\right)^2 \ .
\end{equation}
If either of the scalar fields is fixed, $\epsilon_\sigma$ or
$\epsilon_\phi$ identically zero, then we recover the single field
results where $\zeta$ is constant and equal to $Q_1$ or $Q_3$,
respectively. We see from Eq.~(\ref{DZETA}) that $\zeta$ continues to
evolve after horizon crossing if both fields are evolving and
$\Upsilon \neq 0 $.  In the slow-roll approximation we have
\begin{equation}
\dot\zeta \simeq e^{2(h_*-h)}\,\left(e^{-2h_*}\,Q_3 -
Q_1\right)\,\Upsilon\,.
\end{equation}
However, if $h$ grows significantly during inflation, then the change
in $\zeta$ may be small, as can happen in the case of scalar-tensor
gravity theories~\cite{GBW,DGL}. On the other hand, for minimally
coupled scalar fields, where $h=0$, we have
$\Delta\zeta\simeq(Q_3-Q_1)\int\Upsilon dt$ which is typically of order
$\zeta_*$.

The spectrum of density perturbations at the end of inflation
${\cal P}^e_{\zeta}(k)$ computed from Eq.~(\ref{SRZETA}) is,
\begin{eqnarray}\label{PKZETA}
{\cal P}^e_{\zeta} & \simeq &
{\kappa^2\over2}\,\left({H_*\over2\pi}\right)^2
\left[\left({\epsilon_\phi^e + (e^{2h_e} - e^{2h_*})\,
\epsilon_\sigma^e\over\epsilon_\phi^e
+ e^{2h_e}\, \epsilon_\sigma^e}\right)^2 {1\over\epsilon_\phi^*}
\right. \nonumber \\ [2mm]
& & \ \left. +
\left({\epsilon_\sigma^e\over\epsilon_\phi^e
+ e^{2h_e}\, \epsilon_\sigma^e}\right)^2 {e^{2h_*}
\over\epsilon_\sigma^*}\right] \,,
\end{eqnarray}
where we have used Eqs.~(\ref{Q1}) and~(\ref{Q3}). This is the
result found in Ref.~\cite{GBW}.

We will now show that Eq.~(\ref{PKZETA}) can also be derived in the
framework of Sasaki and Stewart~\cite{SS95} in terms of the change in
the number of $e$-foldings, $\delta N$, as given in
Eq.~(\ref{dNzeta}). We wish to evaluate $\zeta_e$, the curvature
perturbation on a hypersurface of fixed energy density, $U(\phi_e)
V(\sigma_e) =$ constant, near the end of inflation. The values of the
fields $\phi_e$ and $\sigma_e$ at the end of inflation on a given
classical trajectory will be a function of the conserved quantity,
$C$, defined in Eq.~(\ref{sol}).  Therefore,
\begin{equation}\label{aux1}
\left({U'\over U}\right)_e \,{d\phi_e\over dC} +
\left({V'\over V}\right)_e \,{d\sigma_e\over dC} = 0\,.
\end{equation}
Differentiating Eq.~(\ref{sol}) with respect to the trajectory $C$,
and using Eq.~(\ref{aux1}), we find
\begin{eqnarray}\label{aux2}
&{\displaystyle
\kappa^2\,{d\sigma_e\over dC} = \left({V\over V'}\right)_e \,\left[
\left({V\over V'}\right)_e^2 + e^{2h_e}\,\left({U\over U'}
\right)_e^2 \right]^{-1} \,,}\nonumber\\[2mm]
&{\displaystyle
\kappa^2\,{d\phi_e\over dC} = - \left({U\over U'}\right)_e \,\left[
\left({V\over V'}\right)_e^2 + e^{2h_e}\,\left({U\over U'}
\right)_e^2 \right]^{-1} } \,.
\end{eqnarray}
We can now evaluate the dependence of the number of $e$-folds $N$,
Eq.~(\ref{Nefold}), on the initial values $\phi_*$ and $\sigma_*$,
\begin{eqnarray}
dN & = & \kappa^2\,\left({U\over U'}\right)_*\,d\phi_*
\nonumber \\ [2mm]
& & \ - \
\kappa^2\,\left({U\over U'}\right)_e\,{d\phi_e\over dC}\,
\left[{\partial C\over\partial\phi_*}\,d\phi_* +
{\partial C\over\partial\sigma_*}\,d\sigma_* \right] \,.
\end{eqnarray}
Using Eqs.~(\ref{aux2}) and~(\ref{sol}) we find
\begin{eqnarray}
&{\displaystyle
\left|{\partial N\over\partial\phi_*}\right|^2 =
{\kappa^2\over2\epsilon_\phi^*}\,\left[{\epsilon_\phi^e
+ \left(e^{2h_e} - e^{2h_*}\right)\,\epsilon_\sigma^e\over
\epsilon_\phi^e + e^{2h_e}\,\epsilon_\sigma^e}
\right]^2 \,,}\nonumber\\[3mm]
&{\displaystyle
\left|{\partial N\over\partial\sigma_*}\right|^2 =
{\kappa^2\over2\epsilon_\sigma^*}\,\left[{\epsilon_\sigma^e\over
\epsilon_\phi^e + e^{2h_e}\,\epsilon_\sigma^e}\right]^2 \,.}
\end{eqnarray}
These are the expressions required to evaluate ${\cal P}^e_\zeta$ using
Eq.~(\ref{dNzeta}) with $\delta\phi_\ast$ and $\delta\sigma_\ast$
given by Eqs.~(\ref{Pdphi}) and~(\ref{Pdsigma}),
\begin{equation}\label{spectrum}
{\cal P}_\zeta^e =
\left({H_*\over2\pi}\right)^2\,\left[\left|{\partial N\over
\partial\phi_*}\right|^2 + e^{2h_*} \,\left|{\partial N\over
\partial\sigma_*}\right|^2\right] \,.
\end{equation}
This expression exactly coincides with that obtained in
Eq.~(\ref{PKZETA}). Note that this result applies to both
scalar-tensor theories (with $U(\phi) = \exp\{-4h(\phi)\}$, see
Ref.~\cite{GBW}), and for minimally coupled ($h(\phi) = 0$) two-field
inflation with a separable potential in general relativity.

\subsection{Tensor perturbations}

In addition to the scalar curvature perturbations that give rise to
density perturbations, tensor or gravitational wave perturbations
[$h_{ij}$ in Eq.~(\ref{pertbn})] can also be generated from quantum
fluctuations during inflation~\cite{AW83}. These do not couple to the
matter content and are determined only by the dynamics of the
background metric, so the standard results for the evolution of tensor
perturbations of the metric remain valid. The two independent
polarizations evolve like minimally coupled massless fields with a
spectrum~\cite{MFB92,LL93}
\begin{equation}
{\cal P}_g = 8\kappa^2\,\left({H_*\over2\pi}\right)^2 \ .
\end{equation}

Note that the tilt of the gravitational wave spectrum,
$n_g\equiv d\ln{\cal P}_g/d\ln k$, is given by
\begin{equation}\label{NG}
n_g = 2\ {\dot{H} \over H^2}
 \simeq - 2 \left(\epsilon_\phi^* + e^{2h_*}
\epsilon_\sigma^* \right) \,.
\end{equation}
Noting the condition for inflation given in Eq.~(\ref{end}) we see
that the definitive test for inflation is the presence of a
gravitational wave spectrum remains $-2<n_g<0$ and is unaltered by the
presence of more than one scalar field. The actual measurement of this
slope will be exceedingly difficult. Tensor perturbations do not
contribute to structure formation and in many inflationary models the
observable effect of gravitational waves is completely
negligible~\cite{LL93}.

Gravitational wave perturbations can contribute to the microwave
background aniso\-tropies only on the largest scales (scales larger
than the Hubble scale at last-scattering, corresponding to about
$1^\circ$ on the sky).  Their contribution relative to scalar
curvature perturbations is given by the ratio~\cite{LL93}
\begin{equation}
R \simeq {3\over4} {{\cal P}_g \over
 {\cal P}_\zeta} \ .
\end{equation}
The rapid decay of the gravitational wave anisotropies on smaller
scales is their most distinctive signature.  If we define
\begin{equation}\label{G}
G \equiv {\epsilon_\phi\over\epsilon_\sigma} + e^{2h}\, ,
\end{equation}
then we can write the ratio of tensor to scalar contributions as
\begin{equation}\label{R}
R \simeq 12\,\epsilon_\sigma^* \, {G_e^2(G_*-e^{2h_*}) \over
(G_e-e^{2h_*})^2 + e^{2h_*}(G_*-e^{2h_*})} \, .
\end{equation}
Together with Eq.~(\ref{NG}) one can show, see Fig.~1, that
\begin{equation}\label{CR}
R \ \leq \ 6\, |n_g| \ .
\end{equation}
This result was found by Polarski and Starobinsky in their model of
double inflation~\cite{PS95} and given in Ref.~\cite{SS95} for general
multiple-field inflation. It generalizes the usual consistency
relation between $R$ and $n_g$ in single field inflation, where $R
\simeq 6\,|n_g|$, see Ref.~\cite{LL92}. It shows that in principle one
could tell from the spectra of metric perturbations whether more than
one field is evolving during inflation.

The inequality of Eq.~(\ref{CR}) becomes an equality only when $G_* =
G_e$.  Note that $\zeta$ in Eq.~(\ref{SRZETA}) can be written as
\begin{equation}\label{zeta}
\zeta \simeq {(G - e^{2h_*})\,Q_1 + Q_3\over G} \, ,
\end{equation}
and therefore the inequality is saturated whenever $\zeta$ remains
constant after horizon crossing. As shown in Eq.~(\ref{dotzeta}) this
only occurs when the perturbations are adiabatic.  This is, of course,
true when only one field is evolving but we can also find a class of
two-field models in which $G$ is constant in the slow-roll
approximation.

Even perturbations off the classical trajectory are forced to be
adiabatic ($\delta p/\delta\rho=\dot p/\dot\rho$) when the pressure
$p$ is a function solely of the density. In our two-field model, the
pressure and energy density are given by,
\begin{eqnarray}\label{pressure}
p & = &{\displaystyle
{1\over2}\,(\dot\phi^2+e^{-2h}\,\dot\sigma^2) - W(\phi,\sigma)\,, }\\[2mm]
\rho & = &{\displaystyle
{1\over2}\,(\dot\phi^2+e^{-2h}\,\dot\sigma^2) + W(\phi,\sigma)\,, }
\end{eqnarray}
and so $p=p(\rho)$ implies that the kinetic energy density is a
function of the potential. We will now show that in the slow-roll
approximation for a separable potential this leads to $G$ being
conserved. Along the classical trajectories,
\begin{equation}
\dot G = \left({\epsilon_\phi'\over\epsilon_\sigma} + 2\,h'\,e^{2h}
\right)\,\dot\phi - {\epsilon_\phi\,\epsilon_\sigma'\over
\epsilon_\sigma^2}\,\dot\sigma \,.
\end{equation}
If the kinetic energy is to be a function solely of the
potential energy density, we require $\epsilon_\phi + e^{2h}\,
\epsilon_\sigma = f(W)$, which implies that
\begin{equation}
\dot G = {\epsilon_\phi\over\epsilon_\sigma}\,
\left({\dot\phi\over\epsilon_\phi}\,{\partial \ln W\over\partial\phi}
- {\dot\sigma\over e^{2h}\epsilon_\sigma}
  \,{\partial \ln W\over\partial\sigma}
\right) \, {df\over d\ln W}\,,
\end{equation}
which from the slow-roll equations of motion must be zero.
Thus we see that both the relation $R=6\,|n_g|$ and the constancy of
$\zeta$ are a consequence of the scalar perturbations being adiabatic
rather than the number of fields present.

For example, consider the class of generalized Brans-Dicke
theories~\cite{DGG,SY95} where $2h(\phi)=\gamma\kappa\phi$ and
$U(\phi)=e^{-\beta\kappa\phi}$ in the Einstein frame with a polynomial
inflaton potential, $V(\sigma)=\lambda\sigma^{2n}/2n$. In this case,
we have
\begin{equation}
\epsilon_\phi + e^{2h}\,\epsilon_\sigma = {\beta^2\over2} +
e^{\gamma\kappa\phi}\,{2n^2\over\kappa^2\sigma^2} \,.
\end{equation}
We see that for $n=\beta/\gamma$ we have $f(W) = \beta^2/2 + 2n^2/
\kappa^2(2nW/\lambda)^{1/n}$. As discussed above, this ensures that all
perturbations are adiabatic and thus $\zeta$ is conserved. We could
also see this by evaluating $G$, which in this case becomes
$G=\beta\gamma\,C$, where $C$ is the conserved quantity along the
classical trajectory, given in Eq.~(\ref{sol}). Thus $\zeta$ is
constant, see Eq.~(\ref{zeta}), and $R = 6\,|n_g|$.

In particular, we find that $\zeta$ remains a constant after horizon
crossing in Brans-Dicke gravity, where $\beta=2\gamma$, for a quartic
potential $V=\lambda\sigma^4/4$. We will study this case in more
detail in the next subsection.

\subsection{Brans-Dicke case}

We will study here a very simple case which we can solve completely.
This is Brans-Dicke case with a quartic potential for the inflaton
field, $h(\phi) = \alpha\kappa\phi$, $U(\phi) =
e^{-4\alpha\kappa\phi}$, and $V(\sigma) = \lambda\sigma^4/4$.  The
constant $\alpha$ characterizes the relative coupling of scalar and
tensor fields to matter, and is related to the usual Brans-Dicke
parameter by $2\alpha^2 = 1/(2\omega+3)$~\cite{GBW}. As shown in the
preceding section this belongs to the sub-class of two-field models
for which $\zeta$ is in fact conserved on scales outside the horizon
in the slow-roll approximation.

Following Eq.~(\ref{sol}), each classical trajectory can be
para\-metrized by
\begin{equation}\label{BD}
G = 8\alpha^2\,C = {\epsilon_\phi\over\epsilon_\sigma} +
e^{2h} \,,
\end{equation}
where $\epsilon_\phi = 8\alpha^2$ and $\epsilon_\sigma =
8/\kappa^2\sigma^2$ are the slow-roll parameters defined in
Eq.~(\ref{SRP}).

The spectrum of curvature perturbations is then
\begin{eqnarray}\label{spect}
{\cal P}^e_\zeta & \simeq & {\kappa^2\over2}\,\left({H_*\over2\pi}\right)^2\,
{1\over G\,\epsilon_\sigma^*} \, \nonumber \\
& \simeq & {\kappa^2\over2}\,\left({H_*\over2\pi}\right)^2\,
{1\over \epsilon_\phi + e^{2h_*}\epsilon_\sigma^*}
\end{eqnarray}
This is an extremely simple and compact formula which makes it
possible to compute the spectral tilt of the scalar perturbations,
$n_s-1\equiv d\ln{\cal P}^e_\zeta/d\ln k$,
\begin{equation}\label{index}
n_s - 1 = - 2 \epsilon_\phi - 3 e^{2h_*}\, \epsilon_\sigma^*\,.
\end{equation}

The ratio of tensor to scalar perturbations is given by
\begin{equation}
R \simeq 12 \,G\,\epsilon_\sigma^*\,
 \simeq 12 \left(\epsilon_\phi + e^{2h_*}\epsilon_\sigma^* \right) \, .
\end{equation}
Depending on $e^{2h}\epsilon_\sigma$ at horizon crossing, we might
have significant gravitational wave contributions. As shown earlier,
the tensor perturbations' spectral index is given by $n_g\simeq-R/6$,
the same relation as in the single field case.

\subsection{Minimally coupled case}

If we restrict our attention to fields which are minimally coupled,
i.e., $h=0$, with a separable potential, then there is no frictional
damping of the $\sigma$ field and the results simplify
compared with the general case.  The spectrum of curvature
perturbations at the end of inflation in Eq.~(\ref{PKZETA}) becomes
\begin{equation}
{\cal P}^e_{\zeta} \simeq
{\kappa^2\over2}\,\left({H_*\over2\pi}\right)^2
\left[\left({\epsilon_\phi^e \over
 \epsilon_\phi^e + \epsilon_\sigma^e}\right)^2
 {1 \over \epsilon_\phi^*} +
\left({\epsilon_\sigma^e
 \over \epsilon_\phi^e + \epsilon_\sigma^e}\right)^2
 {1 \over \epsilon_\sigma^*}  \right] \, ,
\end{equation}
and the ratio between the tensor and scalar contributions
to the microwave background anisotropies on large scales is given by
\begin{equation}
R \simeq 12 \,{ (\epsilon_\phi^e+\epsilon_\sigma^e)^2
 \ \epsilon_\phi^* \ \epsilon_\sigma^*
 \over
 (\epsilon_\phi^e)^2 \ \epsilon_\sigma^*
 + (\epsilon_\sigma^e)^2 \ \epsilon_\phi^* } \ .
\end{equation}
Only if either $\epsilon_\phi$ or $\epsilon_\sigma$ are zero, or if
both are constants (corresponding to exponential potentials), are all
perturbations adiabatic, $\zeta$ conserved and $R=6\,|n_g|$.

More generally, the perturbations may still become effectively
adiabatic by the end of inflation if the evolution has become
essentially one-dimensional along, say, the $\sigma$ direction in
field-space. This requires not only
$\epsilon_\sigma^e\gg\epsilon_\phi^e$, but also $\zeta$ on a given
scale to be due to perturbations in the $\sigma$ field, which from
Eq.~(\ref{SRZETA}) requires $\epsilon_\sigma^e
\gg\epsilon_\phi^e\,(Q_1/Q_3)\simeq\epsilon_\phi^e\,
\sqrt{\epsilon_\sigma^*/\epsilon_\phi^*}$. In this case $\zeta$ has
become constant by the end of inflation and, just as in the single
field case, we expect it to remain conserved until that scale
re-enters the horizon during radiation- or dust-domination.
Nonetheless we may still see evidence of the second field evolving at
horizon crossing during inflation due to the decrease in the Hubble
rate, $H_*$, caused by the slow-roll of $\phi$ in addition to that due
to the evolution of $\sigma$. The tilt of the scalar
perturbation spectrum in this limit is
\begin{equation}
n_s -1 \simeq - 6\,\epsilon_\sigma^* + 2\,\eta_\sigma^* -
2\,\epsilon_\phi^* \ ,
\end{equation}
where we have introduced a further slow-roll parameter
$\eta_\sigma\equiv V''/\kappa^2V$ which describes the curvature of the
potential along $\sigma$. Only as $\epsilon_\phi^*\to0$ do we recover
the familiar single-field result~\cite{LL92,LL93}.
The tilt of the spectrum of tensor
perturbations is $n_g=-2(\epsilon_\sigma^*+\epsilon_\phi^*)$. Thus the
slope of the second field tends to decrease both $n_s$ and
$n_g$. However the ratio between tensor and scalar modes is given by
$R\simeq12\epsilon_\sigma^*$ and so is a function solely of the slope
of the potential along $\sigma$ at horizon crossing, as in the single
field case. Thus the evolution of the $\phi$ leads to the violation of
the single-field consistency relation, $R\simeq-6n_g$.

\section{Conclusions}

The presence of more than one field evolving during inflation has
important consequences for some of the familiar results quoted in
inflationary cosmology. In the slow-roll approximation, $n$ fields
lead to an $n$-dimensional phase-space. Thus, there is no longer a
single classical trajectory that leads to the end of inflation and
different initial conditions may lead to different end-points.
Quantum fluctuations lead to perturbations not only along the
trajectory but also onto neighboring trajectories. We have entropy as
well as adiabatic perturbations and, in this case, the curvature
perturbation $\zeta$ is no longer conserved on super-horizon scales.
Therefore the perturbation on a given scale is no longer determined
solely by quantities at horizon crossing, but also depends upon the
subsequent evolution. In order to evaluate $\zeta$ at the end of
inflation we must be able to integrate the perturbation along the
trajectory. We have shown how this may be performed for a separable
potential in the slow-roll approximation.

In single-field inflation the ratio between the contribution of tensor
to scalar perturbations of the microwave background on large scales
can be related to the tilt of the gravitational wave spectrum,
$R=-6n_g$~\cite{LL92,recon}. We have shown that this is a consequence
of the scalar perturbations being adiabatic and can also occur, within
the slow-roll approximation, in a limited class of two-field
models. More generally, in two-field models with a separable potential
the ratio obeys the inequality $R\leq6|n_g|$, where $n_g$ is given in
terms of the slow-roll parameters in Eq.~(\ref{NG}). While in
principle this is a distinctive prediction of inflation with more than
one scalar field, it is very difficult to observe. By contrast the
tilt of the scalar spectrum is already constrained by observations but
does not give a model-independent test of multiple-field inflation.

While we have been careful to calculate the evolution of the curvature
perturbation during inflation, we have not attempted to go beyond the
end of inflation. Since the observable quantity is the amplitude of
scalar perturbations at re-entry during the radiation- or
dust-dominated eras, the interpretation of our results is sensitive to
the evolution after inflation and in particular to the dynamics of
reheating. If the end of inflation corresponds to a fixed energy
density then $\zeta$ will be conserved across this hypersurface.  This
is the case when there is a unique end-point and only one field is
evolving at the end of inflation. However in the presence of two or
more fields this need no longer be true and requires a more detailed
study.

An important question remains as to whether, even if a given theory
has many scalar fields, one should expect to see evidence in the
perturbation spectra of more than one field evolving. The evolution
during the final 60 e-foldings of inflation depends upon the initial
conditions, and this may lead to only one field evolving at late
times. For instance, in the case of chaotic inflation in some
scalar-tensor gravity theories, the stochastic evolution may fix the
value of the gravitational coupling, effectively leading to single
field inflation~\cite{JGBDW}. To answer the question of initial
conditions, one needs to understand the stochastic evolution of the
coupled fields at early times~\cite{LLM,GBLL,StochInf}.

\section*{Acknowledgments}

The authors are grateful to Andrew Liddle, David Lyth and Ewan Stewart
for useful discussions. JGB is grateful to the organizers of the
Workshop on Inflation at the Aspen Center for Physics for their
hospitality. JGB and DW were supported by PPARC.



\section*{Figure Captions}
\noindent
Fig.~1. Plot of the ratio $R/6\,|n_g|$ from Eqs.~(\ref{NG})
and~(\ref{R}) as a function of the parameters $a\equiv
G_e\,e^{-2h_*}-1$ and $b\equiv G_*\,e^{-2h_*}-1 =
\epsilon_\phi^*/\epsilon_\sigma^*$.
This ratio reaches a maximum of one for $a=b$, corresponding to
$G_e = G_*$. In the minimally coupled case ($h=0$), we have
$a = \epsilon_\phi^e/\epsilon_\sigma^e$, so both $a$ and $b$ must be
non-negative.


\begin{thebibliography}{99}

\bibitem{Book} A. D. Linde, {\em Particle Physics and Inflationary
	Cosmology}, (Harwood, Chur, Switzerland, 1990).

\bibitem{LL93} A. R. Liddle and D. H. Lyth, Phys. Rep. {\bf 231}, 1
	(1993).

\bibitem{tiltvalue} N. Vittorio, S. Matarrese and F. Lucchin,
	Astrophys. J. {\bf 328}, 69 (1988); A. R. Liddle, D. H. Lyth
	and W. Sutherland, Phys. Lett. {\bf B279}, 244 (1992); R. Cen,
	N. Y. Gnedin, L. A. Kofman and J. P. Ostriker, Astrophys. J.
	Lett. {\bf 399} L11 (1992); J. A. Peacock and S. J. Dodds,
	Mon. Not. Roy. astr. Soc. {\bf 267}, 1020 (1994).

\bibitem{Gorski94} K. M. G\'orski, {\em et al},
	Astrophys. J. Lett. {\bf 430}, L89 (1994).

\bibitem{recon} For a review see J. E. Lidsey, A. R. Liddle, E. W.
	Kolb, E. J. Copeland, T. Barreiro and M. Abney,
	``Reconstructing the Inflaton Potential -- An Overview'',
	preprint astro-ph/9508078 (1995).

\bibitem{GP82} A. H. Guth and S.-Y. Pi, Phys. Rev. Lett. {\bf 49},
	1110 (1982); S. W. Hawking, Phys. Lett. {\bf B115}, 295 (1982);
	A. A. Starobinsky, Phys. Lett. {\bf B117}, 175 (1982).

\bibitem{Hybrid} A. D. Linde, Phys. Lett. {\bf B259}, 38 (1991);
	Phys. Rev. D {\bf 49}, 748 (1994); E. J. Copeland,
	A. R. Liddle, D. H. Lyth, E. D. Stewart and D. Wands,
	Phys. Rev. D {\bf 49}, 6410 (1994).

\bibitem{FirstOrder} A. D. Linde, Phys. Lett. {\bf B249}, 18 (1990);
	F. C. Adams and K. Freese, Phys. Rev. D {\bf 43}, 353 (1991).

\bibitem{ExtInf} D. La and P.J. Steinhardt, Phys. Rev. Lett.
	{\bf 62}, 376 (1989); P.J. Steinhardt and F.S. Accetta,
	Phys. Rev. Lett. {\bf 64}, 2470 (1990); J.D. Barrow and
	K. Maeda, Nucl. Phys. {\bf B341}, 294 (1990);
	J. Garc\'\i a--Bellido and M. Quir\'os, Phys. Lett.
	{\bf B243}, 45 (1990).

\bibitem{Double} A. A. Starobinsky, JETP Lett. {\bf 42}, 152 (1985);
	J. Silk and M. S. Turner, Phys. Rev. D {\bf 35}, 419 (1987);
	D. Polarski and A. A. Starobinsky, Nucl. Phys.
	{\bf B385}, 623 (1992); R. Holman, E. Kolb, S. L. Vadas and Y.
	Wang, Phys. Lett. {\bf B269}, 252 (1991).

\bibitem{inter} L. A. Kofman and A. D. Linde, Nucl. Phys. {\bf B282},
	555 (1987); L. A. Kofman and D. Yu. Pogosyan, Phys. Lett.
	{\bf B214}, 508 (1988); D. S. Salopek, J. R. Bond and J. M.
	Bardeen, Phys. Rev. D {\bf 40}, 1753 (1989).

\bibitem{ExtChaot} A. D. Linde, Phys. Lett. {\bf B238}, 160 (1990).

\bibitem{DGL} N. Deruelle, C. Gundlach and D. Langlois,
	Phys. Rev D{\bf 46}, 5337 (1992).

\bibitem{SY95} A. A. Starobinsky and J. Yokoyama, {\it Density
	fluctuations in Brans-Dicke inflation}, astro-ph/9502002 (1995).

\bibitem{MM92} S. Mollerach and S. Matarrese, Phys. Rev. D{\bf 45},
	1961 (1992).

\bibitem{GBW} J. Garc\'\i a--Bellido and D. Wands,
	{\em Constraints from inflation on scalar-tensor theories},
	preprint SUSSEX-AST-95/6-3, gr-qc/9506050. Phys. Rev. D,
	to appear (1995).

\bibitem{SS95} M. Sasaki and E. D. Stewart, {\em A general analytic
	formula for the spectral index}, preprint astro-ph/9507001
	(1995).

\bibitem{B80} J. M. Bardeen, Phys. Rev. D {\bf 22}, 1882 (1980);
	J. M. Bardeen, P. J. Steinhardt and M. S. Turner, Phys. Rev.
	D{\bf 28}, 679 (1983).

\bibitem{KS84} H. Kodama and M. Sasaki, Prog. Theor. Phys. Supp.
	{\bf 78}, 1 (1984); M. Sasaki, Prog. Theor. Phys. {\bf 76},
	1036 (1986).

\bibitem{MFB92} V. F. Mukhanov, H. A. Feldman and R. H.
	Brandenberger, Phys. Rep. {\bf 215}, 203 (1992).

\bibitem{DM95} N. Deruelle and V. Mukhanov, {\em On matching
	conditions for cosmological perturbations}, preprint
	gr-qc/9503050 (1995).

\bibitem{Lyth85} D. H. Lyth, Phys. Rev. D{\bf 31}, 1792 (1985).

\bibitem{JGBNuclPhys} J. Garc\'\i a-Bellido, Nucl. Phys. {\bf B423},
	221 (1994).

\bibitem{GBLL} J. Garc\'\i a-Bellido, A. D. Linde and D. A. Linde,
	Phys. Rev. D{\bf 50}, 730 (1994).

\bibitem{LPB94} A. R. Liddle, P. Parsons and J. D. Barrow,
	Phys. Rev. D{\bf 50}, 7222 (1994).

\bibitem{AW83} L.F. Abbott and M.B. Wise, Nucl. Phys. {\bf B244},
	541 (1984).

\bibitem{PS95} D. Polarski and A. A. Starobinsky, Phys. Lett.
	{\bf B356}, 196 (1995).

\bibitem{LL92} A. R. Liddle and D. H. Lyth, Phys. Lett. {\bf B 291},
	391 (1992).

\bibitem{DGG} T. Damour, G. Gibbons and C. Gundlach,
	Phys. Rev. Lett. {\bf 64}, 123 (1990);
	Phys. Rev. D{\bf 43}, 3873 (1991).

\bibitem{JGBDW} J. Garc\'\i a--Bellido and D. Wands, {\em General
	relativity as an attractor of scalar-tensor stochastic
	inflation}, preprint SUSSEX-AST-95/3-1, gr-qc/9503049,
	Phys. Rev. D, to appear (1995).

\bibitem{LLM} A. D. Linde, D. A. Linde and A. Mezhlumian,
	Phys. Rev. D {\bf 49}, 1783 (1994).

\bibitem{StochInf} J. Garc\'\i a-Bellido and A. D. Linde,
	Phys. Rev. D {\bf 51}, 429 (1995); J. Garc\'\i a-Bellido
	and A. D. Linde, {\em Stationary solutions in Brans-Dicke
	stochastic inflationary cosmology}, preprint SU-ITP-95/8,
	gr-qc/9504022, Phys. Rev. D, to appear (1995).




\end{thebibliography}
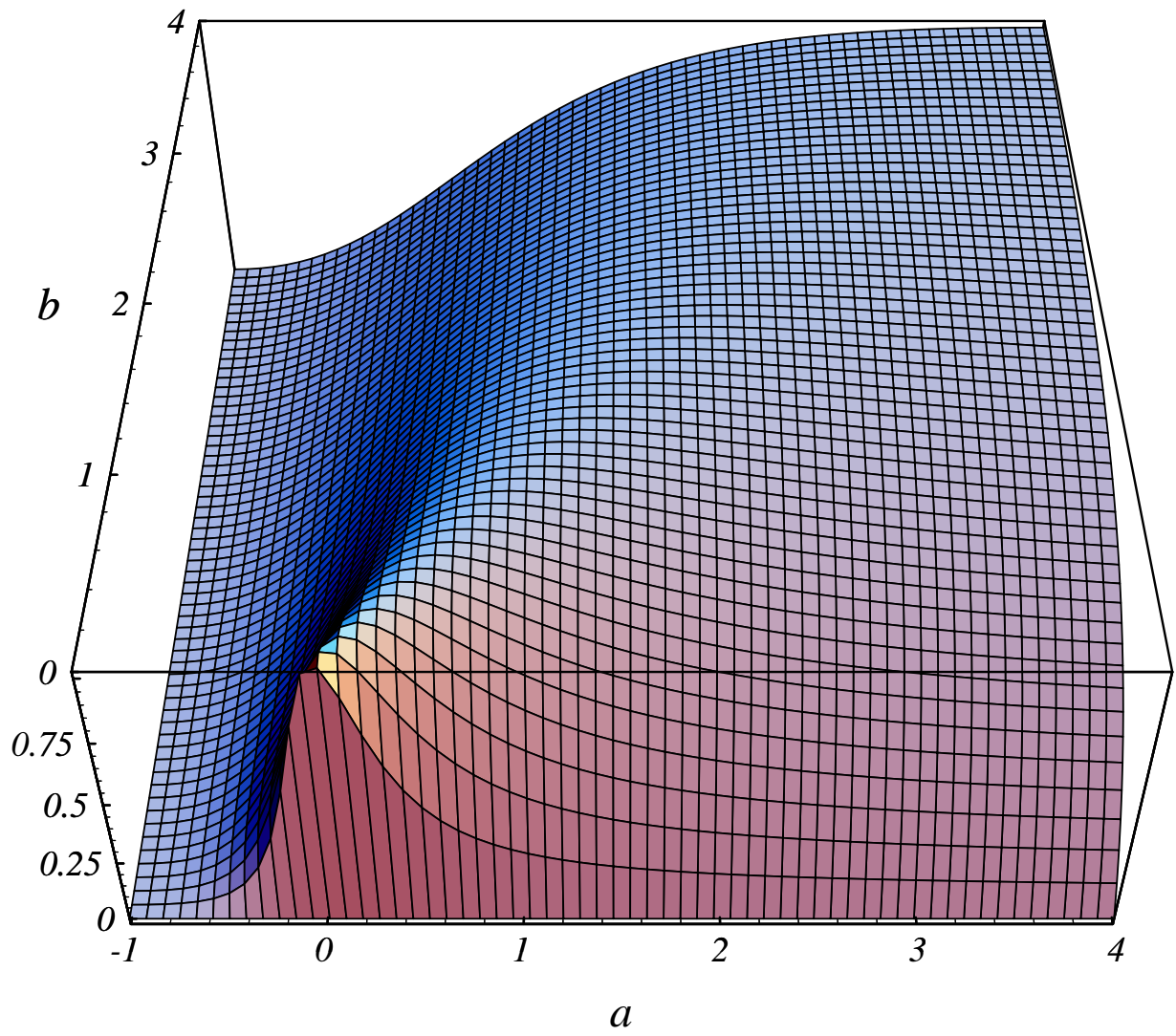
\end{document}